\documentclass[entropy,article,submit,moreauthors,pdftex]{Definitions/mdpi}

\usepackage{graphicx}
\usepackage{amsmath,amssymb}
\renewcommand{\thefootnote}

\newcommand{\beq} {\begin{equation}}
\newcommand{\enq} {\end{equation}}
\newcommand{\ber} {\begin {eqnarray}}
\newcommand{\enr} {\end {eqnarray}}

\newcommand {\ern}[1] {equation (\ref{#1})}

\newcommand {\Sc} {Schr\"{o}dinger}
\newcommand {\SE} {Schr\"{o}dinger equation }

\newcommand{\ce}  {continuity equation }

\firstpage{1}
\pubvolume{1}
\issuenum{1}
\articlenumber{0}
\pubyear{2022}
\copyrightyear{2022}
\externaleditor{Academic Editor: }
\datereceived{}
\dateaccepted{}
\datepublished{}
\hreflink{https://doi.org/}
\Title{Pauli's Electron in Ehrenfest and Bohm Theories, a Comparative Study}
\TitleCitation{Pauli's Electron in Ehrenfest and Bohm Theories, a Comparative Study}

\Author{Asher Yahalom}
\AuthorNames{Asher Yahalom}
\AuthorCitation{Yahalom, A.}

\address{$^{1}$  Department of Electrical \& Electronic Engineering, Faculty of Engineering, Ariel University, Ariel 40700, Israel; asya@ariel.ac.il; Tel.: +972-54-7740294\\
$^{2}$  Center for Astrophysics, Geophysics, and Space Sciences (AGASS), Ariel University,
Ariel 40700, Israel}

\abstract{Electrons moving at slow speeds much lower that the speed of light are described by a wave function which is a solution of Pauli's equation. This is a low velocity limit of the relativistic
Dirac equation. Here we compare two approaches, one which is the more conservative Copenhagen's interpretation denying a trajectory of the electron but allowing a trajectory to the electron expectation value through Ehrenfest theorem. The said expectation value is of course calculated using a solution of Pauli's equation. A less orthodox approach is championed by Bohm, and attributes a velocity field to the electron also derived from the Pauli wave function. It is thus interesting to compare the trajectory followed by the electron according to Bohm and its expectation value according to Ehrenfest. Both similarities and differences will be considered.}

\begin{document}
\nolinenumbers
\section {Introduction}

Quantum mechanics is usually interpreted by the Copenhagen approach. This approach objects to the physical reality of the quantum wave function and declares it to be epistemological (a tool for estimating probability of measurements) in accordance with
the Kantian \cite{Kant} depiction of reality, and its denial of the human ability to grasp any thing in its reality (ontology). However, we also see the development of another approach of prominent scholars that think about quantum mechanics differently.  This school believes in the ontological existence of the wave function. According to this approach the wave function is an element of reality much like an electromagnetic field. This was supported by Einstein and Bohm \cite{Bohm,Holland,DuTe} has resulted in different understandings of quantum mechanics among them the fluid realization championed by Madelung \cite{Madelung,Complex} which stated that the modulus square  of the wave function is a fluid density and the phase is a potential of the velocity field of the fluid.

A non relativistic quantum equation for a spinor was first introduced by Wolfgang Pauli in 1927 \cite{Pauli}, this was motivated by the need to explain the Stern-Gerlach experiments.
Later it was shown that the Pauli equation is a low velocity limit of the relativistic Dirac equation.
This equation is based on a two dimensional operator matrix Hamiltonian. Two dimensional operator matrix Hamiltonians are common in the literature (\cite{EYB1} - \cite{EY8}) and describe many types of quantum systems. A Bohmian analysis of the Pauli equation was given by Holland and others \cite{Holland,DuTe}, however, the analogy of the Pauli theory to fluid dynamics and the notion of spin vorticity were not considered. In \cite{Spflu} spin fluid dynamics was introduces for a single electron with a spin. One thus must contemplate where do those internal energies originate? The answer to this question seems to come from measurement theory \cite{Fisher,Fisherspin}. Fisher information is a basic notion of measurement theory,  and is a figure of merit of a measurement quality of any quantity. It was shown \cite{Fisherspin} that this notion is the internal energy of a spin less electron (up to a proportionality constant) and can be used to partially interpret the internal energy of an electron with spin. An attempt to derive most physical theories from Fisher information is due to Frieden \cite{Frieden}. It was suggested \cite{Fisherspin2} that there exist a velocity field such that the Fisher information will given a complete explanation for the spin fluid internal energy. It was also suggested that one may define comoving scalar fields as in ideal fluid mechanics, however, this was only demonstrated implicitly but not explicitly. A common feature of previous work on the fluid \& Fisher information interpretation of quantum mechanics, is the negligence of electromagnetic interaction thus setting the vector potential to zero, this was recently corrected in \cite{FisherspinElectMag}.

Ehrenfest \cite{Ehrenfest} published his paper in 1927 as well with the title: "Remark on the approximate validity of classical mechanics within quantum mechanics". Using this approach we can accept the orthodox Copenhagen's interpretation denying a trajectory of the electron but at the same time accept the existence of a trajectory of the electron's position vector expectation value through Ehrenfest theorem. The Ehrenfest approach is thus independent of interpretation, and can
be applied according to both the Copenhagen and Bohm schools. However, only in the Bohm approach one
may compare the trajectory of the electron to that of its expectation value.

 We will begin this paper by reminding the reader the basic equation describing the motion
 of a classical electron. This will be followed by a discussion of
 \SE with a non trivial vector potential and its interpretation in terms of Bohmian equation of motion with a quantum force correction. Then we introduce Pauli's equation
 with a vector potential and interpret it in terms a Bohmian equation of motion with a quantum force correction which is different from the \Sc~ case . Finally we derive an equation for Pauli's electron position vector expectation value using Ehrenfest theorem and compare the result to
 the results obtained in Bohm's approach, similarities and differences will arise, a concluding section will follow discussing the Stern-Gerlach experiment.

\section{A Classical Charged Particle}

 Consider a classical particle with the coordinates $\vec x (t)$, mass $m$ and charge $e$ interacting with a given electromagnetic vector potential $\vec A (\vec x,t)$ and scalar potential $\varphi (\vec x,t)$.
 We will not be interested in the effects of the particle on the field and thus consider the field as ``external''. The action of said particle is:
 \ber
 {\cal A} &=& \int_{t1}^{t2} L dt, \qquad L = L_0 + L_i
 \nonumber \\
 L_0 &\equiv& \frac{1}{2} m v^2, \qquad L_i \equiv e(\vec A \cdot \vec v - \varphi), \qquad
  \vec v \equiv \frac{d \vec x}{dt} \equiv \dot{\vec x}, \quad v =|\vec v|.
 \label{classparticleact}
 \enr
The variation of the two parts of the Lagrangian are given by:
\beq
 \delta L_0 =  m \dot{\vec x  } \cdot \delta \dot{\vec x} =
 \frac{d (m \vec v \cdot \delta \vec x)}{dt} - m \dot {\vec v} \cdot \delta \vec x
 \label{delLf}
 \enq
\ber
 \delta L_i &=&  e \left( \delta \vec A \cdot \vec v + \vec A \cdot \delta \dot{\vec x} -
 \delta \varphi \right)
 \nonumber \\
 &=& e \left( \partial_k \vec A \cdot \vec v \delta x_k +
  \frac{d (\vec A \cdot \delta \vec x)}{dt} - \delta \vec x \cdot \frac{d \vec A}{dt}
  - \vec \nabla \varphi \cdot \delta \vec x\right),
 \label{delLi}
 \enr
in the above $\partial_k \equiv \frac{\partial }{\partial x_k}$ and
$\vec \nabla \equiv (\frac{\partial }{\partial x},\frac{\partial }{\partial y},\frac{\partial }{\partial z}) \equiv (\frac{\partial }{\partial x_1},\frac{\partial }{\partial x_2},\frac{\partial }{\partial x_3})$.
We use the Einstein summation convention in which a Latin index (say $k,l$) takes one of the values $k,l \in [1,2,3]$. We may write the total time derivative of $\vec A$ as:
\beq
\frac{d \vec A (\vec x (t),t)}{dt} =  \partial_t \vec A + v_l \partial_l \vec A,
\qquad  \partial_t \equiv \frac{\partial }{\partial t}.
\label{Ader}
\enq
Thus, the variation $\delta L_i$ can be written in the following form:
\beq
 \delta L_i
 =\frac{d (e \vec A \cdot \delta \vec x)}{dt} +
  e \left[ \left(\partial_k  A_l -\partial_l  A_k \right) v_l -  \partial_t A_k
    - \partial_k \varphi \right] \delta x_k,
 \label{delLi2}
 \enq
Defining the electric and magnetic fields in the standard way:
\beq
\vec B \equiv \vec \nabla \times \vec A, \qquad \vec E \equiv  -\partial_t \vec A - \vec \nabla \varphi,
\label{EB}
 \enq
it follows that:
 \beq
\epsilon_{kln} B_n = \partial_k  A_l -\partial_l  A_k  , \qquad  E_k =  -\partial_t  A_k - \partial_k \varphi,
\label{EB2}
 \enq
in which $ \epsilon_{kln}$ is the three index antisymmetric tensor. Thus, we may write $\delta L_i$ as:
 \beq
 \delta L_i
 =\frac{d (e \vec A \cdot \delta \vec x)}{dt} +
  e \left[ \epsilon_{kln} v_l B_n + E_k \right] \delta x_k
  = \frac{d (e \vec A \cdot \delta \vec x)}{dt} +
  e \left[ \vec v \times \vec B + \vec E\right] \cdot \delta \vec x.
 \label{delLi3}
 \enq
 We use the standard definition of the Lorentz force (MKS units):
 \beq
 \vec F_L \equiv e \left[ \vec v \times \vec B + \vec E\right]
 \label{Lor}
 \enq
 to write:
 \beq
 \delta L_i   = \frac{d (e \vec A \cdot \delta \vec x)}{dt} + \vec F_L \cdot \delta \vec x.
 \label{delLi4}
 \enq
 Combining the variation of $L_i$ given in \ern{delLi4} and the variation of $L_0$ given in \ern{delLf}, it follows from \ern{classparticleact} that the variation of $L$ is:
  \beq
  \delta L =\delta L_0 + \delta L_i =
   \frac{d \left( (m \vec v + e \vec A) \cdot \delta \vec x\right)}{dt} +(-m \dot{\vec v} +\vec F_L) \cdot \delta \vec x.
 \label{delL1}
 \enq
 Thus, the variation of the action is:
  \beq
  \delta A =\int_{t1}^{t2} \delta L dt =
  \left. (m \vec v + e \vec A) \cdot \delta \vec x \right|_{t1}^{t2}
   - \int_{t1}^{t2}(m \dot{\vec v} - \vec F_L) \cdot \delta \vec x dt.
 \label{delA1}
 \enq
 Since the classical trajectory is such that the variation of the action on it vanishes for
 a small modification of the trajectory $\delta \vec x$ that vanishes at $t1$ and $t2$ but is otherwise arbitrary it follows that:
 \beq
  m \dot{\vec v} = \vec F_L = e \left[\vec v \times \vec B + \vec E \right]
  \Rightarrow \dot{\vec v} = \frac{e}{m} \left[\vec v \times \vec B + \vec E \right].
 \label{equamotion}
 \enq
 Thus, the dynamics of a classical particle in a given electric and magnetic field is described by a single number, the ratio between its charge and mass:
  \beq
  k \equiv \frac{e}{m} \qquad \Rightarrow \qquad \dot{\vec v} = k \left[\vec v \times \vec B + \vec E \right].
 \label{equamotion2}
 \enq
The reader is reminded that the connection between the electromagnetic potentials and the fields is not unique. Indeed performing  a gauge transformation to obtain a new set of~potentials:
\beq
  \vec A' = \vec A + \vec \nabla \Lambda, \qquad \varphi' = \varphi - \partial_t \Lambda.
 \label{gauge}
 \enq
we obtain the same fields:
\beq
 \vec B' = \vec \nabla \times \vec A' = \vec \nabla \times \vec A = \vec B, \qquad
 \vec E' =   -\partial_t \vec A' - \vec \nabla \varphi' =  -\partial_t \vec A - \vec \nabla \varphi = \vec E.
 \label{gauge2}
 \enq

\section {Schr\"{o}dinger's Theory}

Quantum mechanics according to the Copenhagen interpretation has lost faith in our ability to predict precisely the whereabouts of even a single particle. What the theory does predict precisely is the evolution in time of a quantity
denoted "the quantum wave function", which is related to a quantum particle whereabouts in a statistical manner. This evolution is described by an equation suggested by
Schr\"{o}dinger \cite{Schrodinger}:
\beq
i \hbar \dot{\psi} = \hat{H}_S \psi, \qquad \hat{H}_S = -\frac{1}{2m}
 \left(\hbar \vec \nabla - i e \vec A\right)^2 + e \varphi
 \label{Seq}
 \enq
in the above $i=\sqrt{-1}$ and $\psi$ is the complex wave function. $\dot{\psi}=\frac{\partial \psi }{\partial t}$ is the partial time
derivative of the wave function. $\hbar=\frac{h}{2 \pi}$ is Planck's constant divided by $2 \pi$  and $m$ is the particles mass.
 However, this presentation of quantum mechanics is
 rather abstract and does not give any physical picture regarding the meaning of the quantities involved. Thus we write the quantum wave function using its modulus $a$  and phase $\phi$:
 \beq
\psi = a e^{i\phi }.
\label{psi1}
\enq
 We define the velocity field:
\beq
\vec v_S = \frac{\hbar}{m} \vec \nabla \phi - \frac{e}{m} \vec A
\label{vS}
\enq
 and the mass density is defined as:
 \beq
 \hat \rho= m a^2.
 \label{massden}
 \enq
It is easy to show from \ern{Seq} that the \ce is satisfied:
 \beq
 \frac{ \partial\hat \rho }{ \partial t} + \vec \nabla \cdot (\hat \rho \vec v_S) = 0
 \label{nrec}
 \enq
  Hence $\vec v_S$ field is the velocity associated with  mass conservation.
However, it is also the mass associate with probability $a^2$ (by Born's
 interpretational postulate) and charge density $\rho= e a^2$.
 The equation for the phase $\phi$ derived from \ern{Seq} is as follows:
 \beq
  \hbar \frac{\partial \phi}{\partial t} +
  \frac{1}{2m}\left(\hbar \vec \nabla \phi -e \vec A\right)^2 +  e \varphi =  \frac{\hbar^2 \nabla^2 a }{ 2 m a}
 \label{nrhje}
 \enq
In term of the velocity defined in \ern{vS} one obtains the following equation of motion
(see Madelung \cite{Madelung} and Holland \cite{Holland}):
\beq
\frac{d \vec v_S}{d t}  = \frac{\partial \vec v_S}{\partial t} + (\vec v_S \cdot \vec \nabla) \vec v_S
= - \vec \nabla \frac{Q}{m}+k (\vec E + \vec v \times \vec B)
\label{EulerS}
\enq
 The right hand side of the above equation contains the "quantum correction":
 \beq
Q = -\frac{\hbar^2}{2 m} \frac{\vec{\nabla}^2 \sqrt{\hat \rho}}{\sqrt{\hat \rho}}.
\label{qupo}
\enq
For the meaning of this correction in terms of information theory see:
\cite{Spflu,Fisherspin,Fisherspin2}.  These results  illustrates the advantages
 of using the two variables, phase and modulus, to obtain equations of motion
that have a  substantially different form than the familiar \SE (although having the same
mathematical content) and have straightforward physical interpretations \cite{Bohm}.

The quantum correction $Q$ will of course disappear in the classical limit $\hbar \rightarrow 0$, but even if one intends to consider the quantum equation in its full rigor, one needs to take into
account the expansion of an unconfined wave function. As $Q$ is related to the typical gradient
of the wave function amplitude it follows that as the function becomes smeared over time and the gradient becomes small the quantum correction becomes negligible. To put in quantitative terms:
\beq
\vec F_Q = - \vec \nabla Q \simeq \frac{\hbar^2}{2 m L_R^3}, \qquad
  L_R \simeq \frac{R}{|\vec \nabla R|}
\label{quantcorrect}
\enq
in which $L_R$ is the typical length of the amplitudes gradient. Thus:
 \beq
 |F_Q|  << |F_L| \Rightarrow L_R >> L_{Rc} = \left(\frac{\hbar^2}{2 m F_L}\right)^\frac{1}{3}.
\label{quantcorrect2}
\enq
in which $\vec F_L$ is the classical Lorentz force given in \ern{Lor}. For the current
application in which an electron transverses many rotations of macroscopic size this terms in not important.

\section {Pauli's Theory}

\Sc's quantum mechanics is limited to the description of spin less particles. Indeed the need for
spin became necessary as \SE could not account for the result of the Stern Gerlach experiments, predicting a single spot instead of the two spots obtained for hydrogen atoms. Thus Pauli introduced his equation for a non-relativistic particle with spin is given by:
\beq
i \hbar \dot{\psi} = \hat{H}_P \psi, \qquad \hat{H}_P = -\frac{\hbar^2}{2m}[\vec \nabla-\frac{ie}{\hbar }\vec A]^2 + \mu \vec B \cdot \vec \sigma + e \varphi
= \hat{H}_S ~ I + \mu \vec B \cdot \vec \sigma
 \label{Pauli}
 \enq
$\psi$ here is a two dimensional complex column vector (also denoted as spinor), $\hat{H}_P$ is a two dimensional hermitian operator matrix, $\mu$ is the magnetic moment of the particle, and $I$ is
a two dimensional unit matrix. $\vec \sigma$ is a vector of two dimensional Pauli matrices which can be represented as follows:
\beq
\sigma_{1} = \left( \begin{array}{cc} 0 & 1 \\ 1 & 0 \end{array} \right), \qquad
\sigma_{2} = \left( \begin{array}{cc} 0 & -i \\ i & 0 \end{array} \right), \qquad
\sigma_{3} = \left( \begin{array}{cc} 1 & 0 \\ 0 & -1 \end{array} \right).
\label{sigma}
\enq
The ad-hoc nature of this equation was later amended as it became clear that this is the non relativistic limit of the relativistic Dirac equation.
A spinor $\psi$ satisfying \ern{Pauli} must also satisfy a continuity equation of the form:
\beq
\frac{\partial{\rho_p}}{\partial t} + \vec \nabla \cdot  \vec j  = 0.
\label{massconp}
\enq
In the above:
\beq
\rho_p =  \psi^\dagger \psi, \qquad \vec j =  \frac{\hbar}{2mi} [\psi^\dagger \vec \nabla \psi -(\vec \nabla \psi^\dagger) \psi]
- k \vec{A} \rho_p.
\label{pcd}
\enq
The symbol $\psi^\dagger$ represents a row spinor (the transpose) whose components are equal to the complex conjugate of
the column spinor $\psi$. Comparing the standard continuity equation to \ern{massconp} suggests the definition of a velocity field as follows \cite{Holland}:
\beq
\vec v = \frac{\vec j}{\rho_p}= \frac{\hbar}{2mi\rho_p} [\psi^\dagger \vec \nabla \psi -(\vec \nabla \psi^\dagger) \psi] - k \vec{A}.
\label{pv}
\enq
 Holland \cite{Holland} has suggested the following representation of the spinor:
 \beq
\psi = R e^{i\frac{\chi}{2}} \left( \begin{array}{c} \cos \left(\frac{\theta}{2}\right) e^{i\frac{\phi}{2}} \\
i \sin \left(\frac{\theta}{2}\right) e^{-i\frac{\phi}{2}}  \end{array} \right) \equiv
\left( \begin{array}{c} \psi_{\uparrow} \\ \psi_{\downarrow} \end{array} \right).
\label{psiH}
\enq
In terms of this representation the density is given as:
\beq
R^2 = \psi^\dagger \psi = \rho_p  \Rightarrow R= \sqrt{\rho_p}.
\label{rhopsi}
\enq
The mass density is given as:
\beq
\hat \rho = m \psi^\dagger \psi =m R^2 = m \rho_p .
\label{rhom}
\enq
The probability amplitudes for spin up and spin down electrons are given by:
\beq
a_{\uparrow} = \left | \psi_{\uparrow} \right| = R \left | \cos \frac{\theta}{2} \right|, \qquad
a_{\downarrow} = \left | \psi_{\downarrow} \right| = R \left | \sin \frac{\theta}{2} \right|
\label{probamp}
\enq
Let us now look at the expectation value of the spin:
\beq
<\frac{\hbar}{2} \vec \sigma> = \frac{\hbar}{2}\int \psi^\dagger \vec \sigma \psi d^3 x =
 \frac{\hbar}{2}\int \left(\frac{\psi^\dagger \vec \sigma \psi}{\rho_p}\right) \rho_p d^3 x
\label{spinex}
\enq
The spin density can be calculated using the representation given in \ern{psiH} as:
\beq
\hat s \equiv \frac{\psi^\dagger \vec \sigma \psi}{\rho_p} = (\sin \theta \sin \phi, \sin \theta \cos \phi, \cos \theta), \qquad |\hat s| = \sqrt{\hat s \cdot \hat s} = 1.
\label{spinden}
\enq
This gives an easy physical interpretation to the variables $\theta,\phi$ as angles which describe the projection
of the spin density on the axes. $\theta$ is the elevation angle of the spin density vector and $\phi$ is
the azimuthal angle of the same. The velocity field can now be calculated by inserting $\psi$ given in \ern{psiH} into \ern{pv}:
\beq
\vec v = \frac{\hbar}{2m} (\vec \nabla \chi + \cos \theta \vec \nabla \phi) - k \vec{A}.
\label{pv2}
\enq
We are now in a position to calculate the material derivative of the velocity and obtain the equation of motion for a particle with  (\cite{Holland} p. 393 equation (9.3.19)):
\beq
\frac{d \vec v}{d t}  = - \vec \nabla ( \frac{Q}{m})-\left(\frac{\hbar}{2m}\right)^2
\frac{1}{\rho_p} \partial_k(\rho_p \vec \nabla \hat s_j \partial_k \hat s_j)
+k (\vec E + \vec v \times \vec B) - \frac{\mu}{m} (\vec \nabla B_j) \hat s_j.
\label{EulerP}
\enq
The Pauli equation of motion differs from the classical equation motion and the \Sc~ equation of motion. In addition to the \Sc~ quantum force correction we have an additional spin quantum force correction:
\beq
\vec F_{QS} \equiv  -\frac{\hbar^2}{4 m}
\frac{1}{\rho_p} \partial_k(\rho_p \vec \nabla \hat s_j \partial_k \hat s_j)
= -\frac{\hbar^2}{4 m} \left[
\partial_k( \vec \nabla \hat s_j \partial_k \hat s_j)
+ \frac{\partial_k \rho_p }{\rho_p} \vec \nabla \hat s_j \partial_k \hat s_j
\right]
\label{FQS}
\enq
as well as a term characterizing the interaction of the spin with a gradient of the magnetic field.
\beq
\vec F_{grad B S} \equiv  - \mu (\vec \nabla B_j) s_j
\label{FgrBS}
\enq
As both the upper and lower spin components of the wave function are expanding in free space the gradients which appear in $\vec F_{QS}$ will tend to diminish for any macroscopic scale making this
force negligible. To estimate the condition qualitatively we introduce the typical spin length:
\beq
L_s = {\rm min}_{~i\in\{1,2,3\}} ~|\vec \nabla \hat s_i|^{-1}
\label{Ls}
\enq
Using the above definition we may estimate the spin quantum force:
\beq
 F_{QS} \approx \frac{\hbar^2}{4 m} [\frac{1}{L_s^3} + \frac{1}{L_s^2 L_R}]
=  \frac{\hbar^2}{4 m L_s^2} [\frac{1}{L_s} + \frac{1}{L_R}]
\label{FQSest}
\enq
this suggested the definition of the hybrid typical length:
\beq
 L_{sR} = [\frac{1}{L_s} + \frac{1}{L_R}]^{-1} = \left\{
                                                          \begin{array}{cc}
                                                            L_s & L_s \ll L_R \\
                                                            L_R & L_R \ll L_s \\
                                                          \end{array}
                                                        \right. .
\label{LSR}
\enq
In terms of this typical length we may write:
\beq
 F_{QS} \approx \frac{\hbar^2}{4 m L_s^2  L_{sR}}
\label{FQSest2}
\enq
Thus the conditions for a classical trajectory become:
 \beq
 F_{QS} \ll F_L  \Rightarrow L_s^2  L_{sR} \gg \frac{\hbar^2}{4 m F}  \Rightarrow
 L_s \gg \left\{
     \begin{array}{cc}
       \left(\frac{\hbar^2}{4 m F_L}\right)^{\frac{1}{3}}  & L_s \ll L_R \\
        \left(\frac{\hbar^2}{4 m F_L L_{R}}\right)^{\frac{1}{2}} & L_R \ll L_s \\
        \end{array}
    \right. .
\label{FQSest3}
\enq
Another important equation derived from \ern{Pauli} is the equation of motion for the spin orientation vector
(\cite{Holland} p. 392 equation (9.3.16)):
\beq
\frac{d \hat s}{d t}  = \frac{2 \mu}{\hbar} \vec B_{eff} \times \hat s,
\qquad
\vec B_{eff} = \vec B  - \frac{\hbar^2}{4 \mu m R^2} \partial_i (\rho \partial_i  \hat s)
\label{Spinequ}
\enq
The quantum correction to the magnetic field explains \cite{Holland} why a spin pick up the orientation of the filed in a Stern-Gerlach experiment instead of precessing around it
as a classical magnetic dipole would.

\section{Ehrenfest Theorem}

The above electron equations of motion are not accepted by all quantum physicists. Physicist who follow the Copenhagen school of quantum mechanics declare that a quantum electron does not have a trajectory. However, all quantum physicists agree that one can describe the trajectory of the expectation value of various operators such as position and momentum associated with the electron's trajectory. This calculation is done through Ehrenfest Theorem \cite{Griffiths}. The theorem states
that for every quantum operator $A_o$ with expectation value:
\beq
<A_o> = \int d^3 x \psi^{\dagger} A_o \psi
\label{expect}
\enq
the following equality holds:
\beq
\frac{d <A_o>}{d t} = < \frac{\partial A_o}{\partial t}> + \frac{1}{i \hbar}<[A_o,\hat H]>,
\quad [A_o,\hat H] \equiv A_o \hat H - \hat H A_o.
\label{Ehren}
\enq
The position and velocity operators defined as \cite{Griffiths}:
\beq
\vec x_o  \equiv  \vec x, \qquad \vec v_o  \equiv  \frac{1}{m} (\vec p_o - e \vec A)
= \frac{1}{m} ( -i \hbar \vec \nabla - e \vec A)
\label{posvelop}
\enq
For a \Sc's electron (that is without spin) the following results are obtained by Griffiths \cite{Griffiths} by inserting the above operators into \ern{Ehren}:
\ber
\frac{d <\vec x_o>}{d t} &=&  \frac{1}{i \hbar}<[\vec x,\hat H_S]> = <\vec v_o>
\nonumber \\
\frac{d <\vec v_o>}{d t} &=&  < \frac{\partial \vec v_o}{\partial t}> + \frac{1}{i \hbar}<[\vec v_o,\hat H_S]>
\nonumber \\
&=& \frac{k}{2}<\vec v_o \times \vec B - \vec B \times \vec v_o> + k <\vec E>
\label{EhrenSc}
\enr
Thus the electron position expectation value satisfies the equation:
\beq
\frac{d^2<\vec x_o>}{d t^2} = \frac{k}{2}<\vec v_o \times \vec B - \vec B \times \vec v_o> + k <\vec E>
\label{EhrenSc2}
\enq
this equation resembles the classical and quantum equations of motion but also differs from them
in many important aspects. First let us compare it with \ern{equamotion}, let us also assume that $\vec B = 0$. In this case:
\beq
\frac{d^2<\vec x>}{d t^2} = k <\vec E (\vec x,t)> \neq k \vec E (<\vec x>,t),
\label{EhrenSc3}
\enq
thus as noted by many authors, even in this case the expectation value equations differ from the classical equation of motion except for a very restrictive class of linear electric fields. The difference is even more pronounced for the case $\vec B \neq 0$ which only takes a conventional "Lorentz force" form for a constant magnetic field $\vec B$ \cite{Griffiths}:
\beq
\frac{d^2<\vec x>}{d t^2} = k (<\vec v_o> \times \vec B +  <\vec E>).
\label{EhrenSc4}
\enq
Comparing \ern{EhrenSc2} with the quantum motion \ern{EulerS}, we see that the expectation value equation does not contain a quantum force term, which is a further justification to our assumption that this term may be neglected on macroscopic scales.

Let us now turn our attention to the more realistic Pauli electron which does posses spin,
\ern{EhrenSc} now take the form:
\ber
& &  \frac{d <\vec x_o>}{d t}  = \frac{1}{i \hbar}<[\vec x,\hat H_P]> = \frac{1}{i \hbar}<[\vec x,\hat H_S]> = <\vec v_o>
\nonumber \\
& &  \frac{d <\vec v_o>}{d t} =  < \frac{\partial \vec v_o}{\partial t}> + \frac{1}{i \hbar}<[\vec v_o,\hat H_P]>
\nonumber \\
& & = < \frac{\partial \vec v_o}{\partial t}> + \frac{1}{i \hbar}<[\vec v_o,\hat H_S]>
+ \frac{1}{i \hbar}<[\vec v_o,\mu \vec B \cdot \vec \sigma]>
\nonumber \\
& & = \frac{k}{2}<\vec v_o \times \vec B - \vec B \times \vec v_o> + k <\vec E>
+ \frac{\mu}{i \hbar}<[\vec v_o, B_i] \sigma_i>,
\label{EhrenP}
\enr
in which in the last term we use the Einstein summation convention. However:
\beq
[\vec v_o, B_i] = \frac{1}{m} [-i \hbar \vec \nabla - e \vec A, B_i] =
-\frac{i \hbar}{m} [ \vec \nabla , B_i]
\label{EhrenP2}
\enq
The value of this commutation relation can be deduced by operating with the above operator on
an arbitrary wave function $\psi$.
\beq
[\vec v_o, B_i] \psi =
-\frac{i \hbar}{m} [ \vec \nabla , B_i] \psi =
-\frac{i \hbar}{m} \left( \vec \nabla (B_i \psi) - B_i \vec \nabla \psi \right) =
-\frac{i \hbar}{m} (\vec \nabla B_i) \psi
\label{EhrenP3}
\enq
Or in operator jargon:
\beq
[\vec v_o, B_i]  = -\frac{i \hbar}{m} (\vec \nabla B_i)
\label{EhrenP4}
\enq
It thus follows that:
\beq
\frac{\mu}{i \hbar}<[\vec v_o, B_i] \sigma_i> =
- \frac{\mu}{m}<(\vec \nabla B_i) \sigma_i> = - \frac{\mu}{m}<(\vec \nabla B_i) \hat s_i>
\label{EhrenP5}
\enq
in which we have used \ern{spinden}. Inserting \ern{EhrenP5} into \ern{EhrenP}:
\beq
\frac{d^2 <\vec x_o>}{d t^2}  =  \frac{d <\vec v_o>}{d t}
 = \frac{k}{2}<\vec v_o \times \vec B - \vec B \times \vec v_o> + k <\vec E>
- \frac{\mu}{m}<\hat s_i \vec \nabla B_i>
\label{EhrenP6}
\enq
comparing the above equation to \ern{EulerP} it follows that the only quantum force surviving
the expectation value averaging is the one describing the effect of the magnetic field gradient on the spin vector which is in accordance with what should be expected in the macroscopic limit
thus leading to the Stern-Gerlach experiment.

\section{Conclusion: Spin orientation and the Stern-Gerlach experiment}

We have seen how the Ehrenfest theorem approach causes "quantum forces" to disappear,
those force disappear also in Bohm's approach if one consider macroscopic scales of propagation.

\begin{figure}
\centering
\includegraphics[width= 0.7\columnwidth]{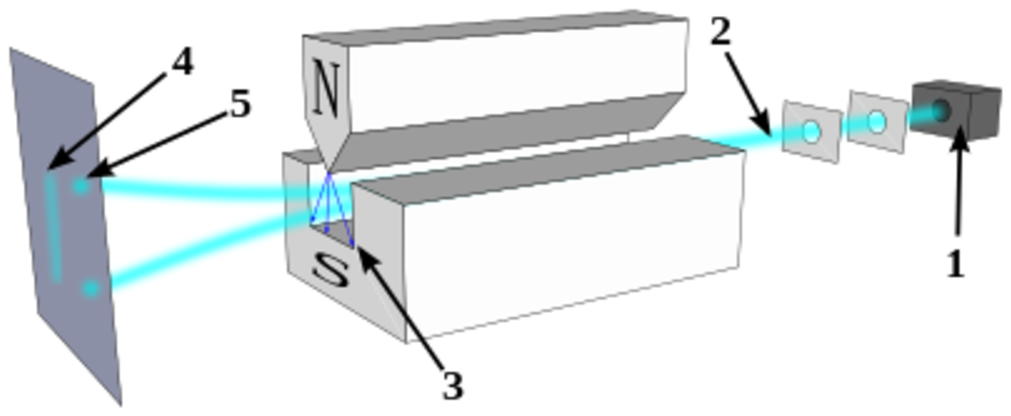}
\caption{A schematics of Stern-Gerlach experiment. Neutral particles travelling through an inhomogeneous magnetic field, and being deflected up or down depending on their spin; (1) particle source, (2) beam of particles, (3) inhomogeneous magnetic field, (4) classically expected result (neglecting the quantum spin force), (5) observed result.}
 \label{SternGerlach}
\end{figure}

\begin{figure}
\centering
\includegraphics[width= 0.7\columnwidth]{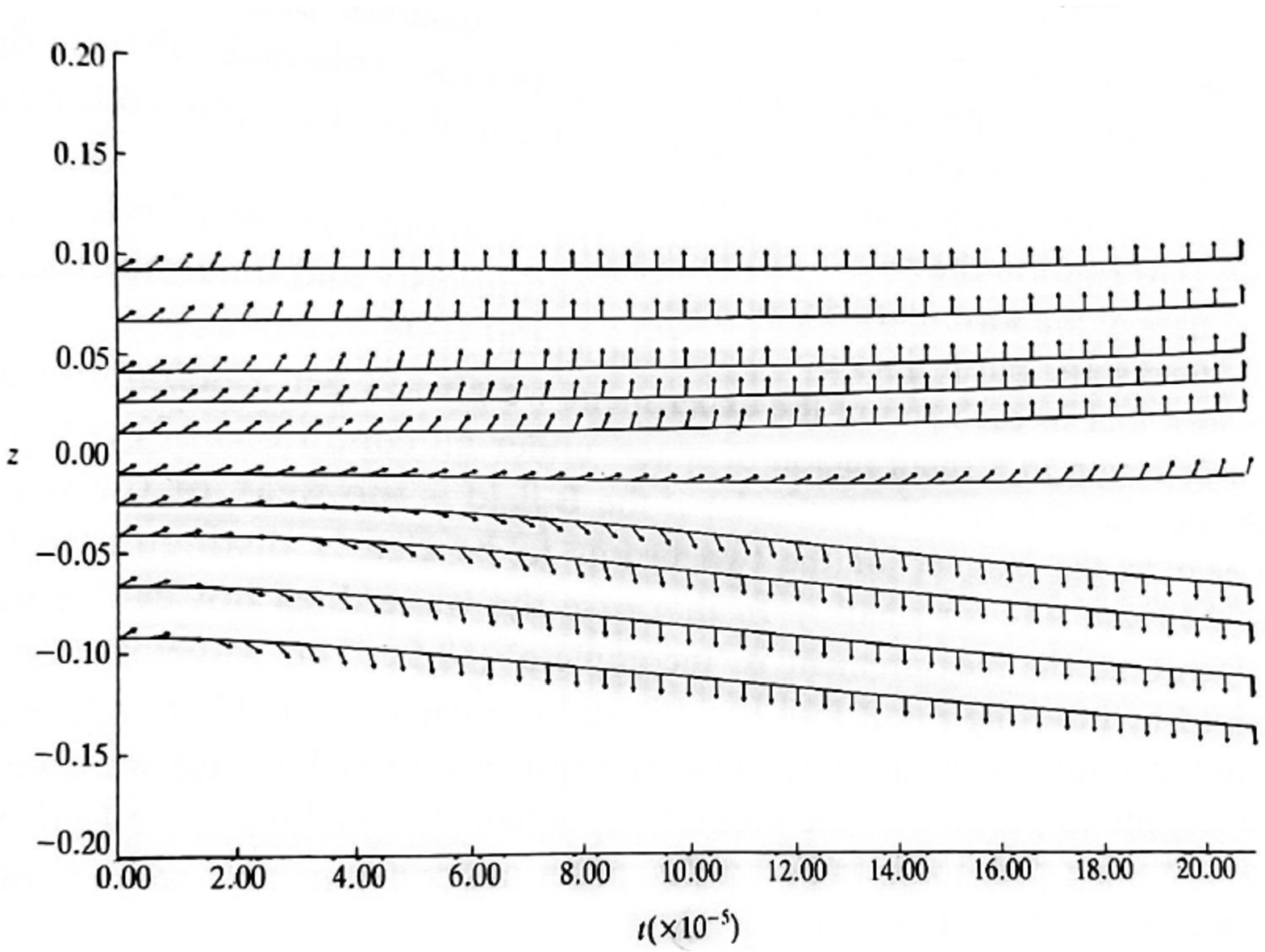}
\caption{A  Stern-Gerlach simulation (Holland \cite{Holland} figure 9.13). Notice how the spin components align in opposite directions in each beam.}
 \label{SternGerlachsim}
\end{figure}

The Stern-Gerlach experiment is an example of using the spin force term given in \ern{FgrBS}
to separate spin up and spin down particles thus obtaining from a single ray of particles two
spots. If the magnetic field  gradient is dominant in a single direction (say $z$) we may write \ern{FgrBS}
as:
\beq
 F_{grad B S~z} =  - \mu (\partial_z B_j) \hat s_j
\label{FgrBS2}
\enq
hence depending on the values $\hat s_j$ some particle will move up and some will move down creating
two spots (see figure \ref{SternGerlach}).
The Stern-Gerlach experiment is usually performed with natural particle not with charged particles like electrons, the reason for this is that generally speaking the classical Lorentz forces are much stronger than the quantum spin force and thus the two spot effect is not observed. Holland shows by simulating \ern{Spinequ} that the spins in a Stern-Gerlach rotate in the direction or opposite to the direction of the magnetic field depending on the trajectory of the particle, that is to which spot it belongs (see figure \ref{SternGerlachsim} (Holland \cite{Holland} figure 9.13)). Notice, however, that from an energy perspective the lowest energy belong to the case in which the spin (and thus its related magnetic dipole) point at the direction of the field.
The energy value is given by the expectation value of the Hamiltonian:
\beq
E = <\hat H_P> = <\hat H_S> + \mu <\vec B \cdot \vec \sigma >.
\label{Ene}
\enq
If the direction of the field $\vec B$ is defined as a $z$ direction it follows that:
\beq
E = <\hat H_S> + \mu < B_z  \sigma_z > =<\hat H_S> + \mu \int d^3 x B_z (a_{\uparrow}^2 -a_{\downarrow}^2)  .
\label{Enez}
\enq
So particles with a definite spin direction (up or down) may have an upper or lower energy
depending on the value of $\mu$. For an electron $\mu$ is the Bohr magneton:
\beq
\mu = \mu_B =\frac{|e| \hbar}{2 m}.
\label{mu}
\enq
and thus lower spin electrons will have a lower energy, if $B_z$ is constant we may write this term
in the "classical" form using a magnetic dipole:
\beq
E = <\hat H_S>  - \vec \mu \cdot \vec B, \qquad
\vec \mu =  -  \mu \int d^3 x (a_{\uparrow}^2 -a_{\downarrow}^2) \hat z  =
\mu \int d^3 x (a_{\downarrow}^2-a_{\uparrow}^2) \hat z.
\label{Enez2}
\enq
hence the magnetic dipole will point in the direction of the field for a lower energy configurations
(spin down) and in the opposite direction for the higher energy configuration (spin up). As systems
tend to relax to their lower energy state, one may ask why do the particles in a Stern-Gerlach experiment do not relax to the spin down configuration and instead split to beams of spin up and spin down with about the same size (see figure \ref{SternGerlachsim})? The answer may be connected
to the fact that in this type of experiment the electrons feel the magnetic field for only a short while and do not have enough time to relax to their minimum energy configurations.

This is not the case in NMR and MRI experiment in which the magnetic dipoles are under the influence of a strong magnetic field, for a long duration. In those cases the magnetization defined as:
\beq
\vec M  = \rho_P \vec \mu
\label{Magnetizaion}
\enq
Satisfies the Bloch phenomenological equations:
\ber
\frac{d M_x}{dt} &=& \gamma_{gyro} (\vec M \times \vec B)_x   - \frac{M_x}{T_2}
\nonumber \\
\frac{d M_y}{dt} &=& \gamma_{gyro} (\vec M \times \vec B)_y   - \frac{M_y}{T_2}
\nonumber \\
\frac{d M_z}{dt} &=& \gamma_{gyro} (\vec M \times \vec B)_z   - \frac{M_z-M_0}{T_1}
\label{Bloch}
\enr
in the above $T_1$ and $T_2$ are typical relaxation times and $\gamma$ is a gyromagnetic ratio
which a for an electron takes the value:
\beq
\gamma_{gyro} = \frac{2 \mu_B}{\hbar}
\label{gyromag}
\enq
The magnetization satisfying the above equation eventually relaxes to the direction of the field which is the minimal energy configuration.
\begin{figure}
\centering
\includegraphics[width= 0.4\columnwidth]{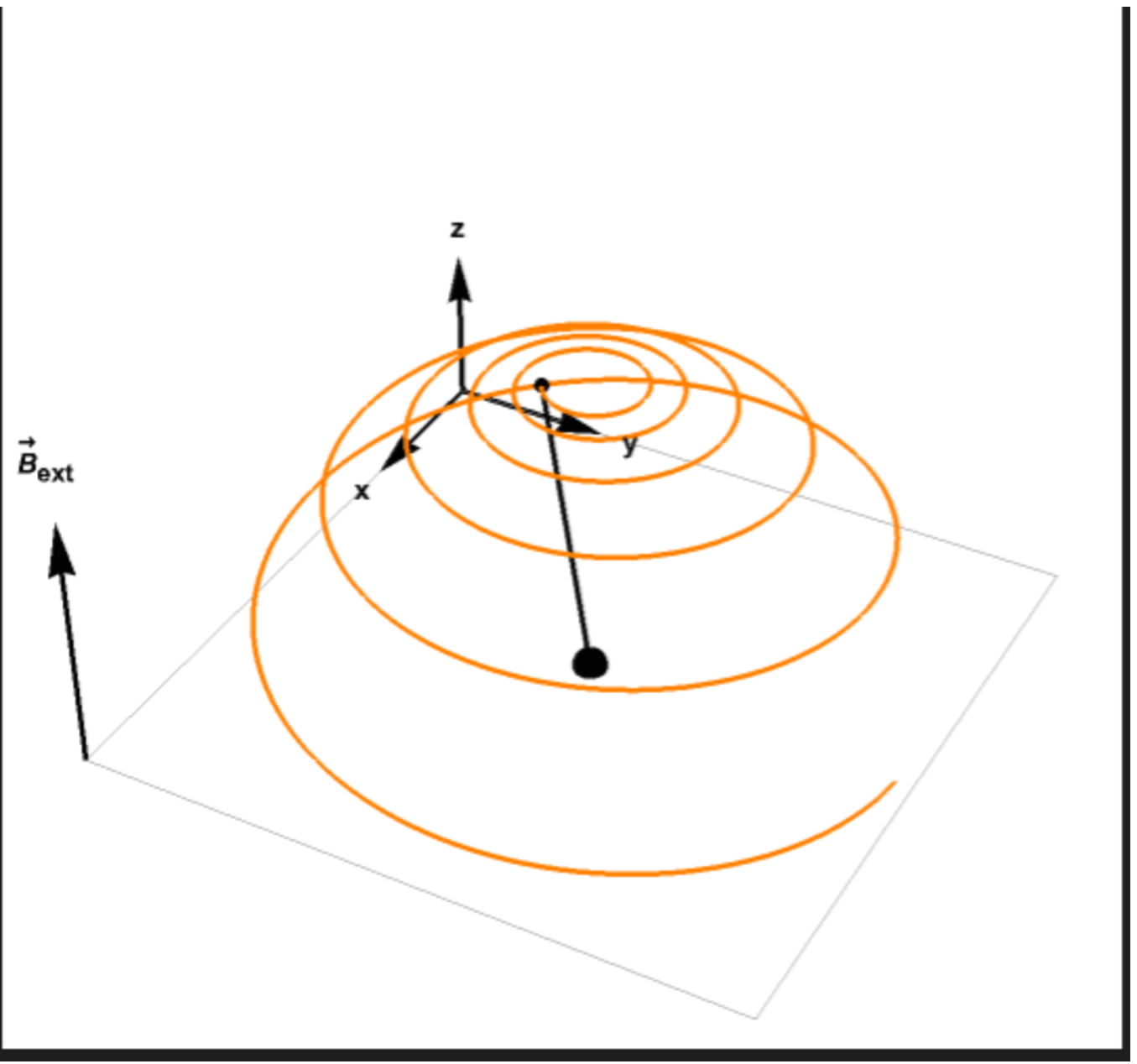}
\caption{The evolution of magnetization towards relaxation, the tip of the magnetization vector
is described by the orange line.}
 \label{Blochevolution}
\end{figure}

As a future direction to the current research it may be interesting to study the same problem
for a fully relativistic electron, however, this will require using a Dirac equation rather than
Pauli's equation.

\end{paracol}
\reftitle{References}

\end {document}